\documentstyle[preprint,aps,epsfig,amssymb]{revtex}
\tighten
\begin{document}
\draft
\preprint{Preprint Numbers: \parbox[t]{45mm}{KSU-CNR-112-99\\
                                             nucl-th/9910033}}

\title{The Quark-Photon Vertex and the Pion Charge Radius}

\author{Pieter Maris and Peter C. Tandy} 
\address{Center for Nuclear Research, Department of Physics,\\ 
         Kent State University, Kent OH 44242}
\date{\today}
\maketitle
\begin{abstract}
The rainbow truncation of the quark Dyson--Schwinger equation is
combined with the ladder Bethe--Salpeter equation for the dressed
quark-photon vertex to study the low-momentum behavior of the pion
electromagnetic form factor.  With model gluon parameters previously
fixed by the pion mass and decay constant, the pion charge radius
$r_\pi$ is found to be in excellent agreement with the data.  When the
often-used Ball--Chiu Ansatz is used to construct the quark-photon
vertex directly from the quark propagator, less than half of $r_\pi^2$
is generated.  The remainder of $r^2_\pi$ is seen to be attributable to
the presence of the $\rho$-pole in the solution of the ladder
Bethe--Salpeter equation.
\end{abstract}

%

\section{Introduction}

For timelike photon momenta $Q^2$ in the vicinity of the $\rho$-meson
mass-shell, the pion charge form factor $F_\pi(Q^2)$ will exhibit a
resonant peak associated with the propagation of intermediate state
$\rho$-mesons (we ignore a possible small effect due to $\rho-\omega$
mixing).  That is
\begin{equation}
 {}F_\pi(Q^2) \to \frac{ g_{\rho\pi\pi} \; m_\rho^2} 
        { g_\rho \,  (Q^2 + m_\rho^2 - im_\rho \, \Gamma_\rho) } \;, 
 \label{rhopole}
\end{equation}
in the Euclidean metric ($Q^2<0$ corresponds to the timelike region)
that we use throughout this work.  Here \mbox{$m_\rho^2/g_\rho$} is the
\mbox{$\rho-\gamma$} coupling strength fixed by the \mbox{$\rho
\rightarrow e^+ \, e^-$} decay, $g_{\rho \pi\pi}$ is the coupling 
constant for the \mbox{$\rho \rightarrow \pi \,\pi$} decay, and
$\Gamma_\rho$ is the $\rho$ width which is principally due to the latter
process.  A long-standing issue in hadronic physics is the question of
the extent to which $F_\pi(Q^2)$ at low spacelike $Q^2$ can be described
by the $\rho$-resonance mechanism.  This is an essential element of the
vector meson dominance (VMD) model which is one of the earliest field
theory models to be successful in a point coupling description of
aspects of hadron dynamics.  In the form of VMD where $\rho-\gamma$
coupling is described by the contraction of the two field strength
tensors $\rho^{\mu \nu}F_{\mu \nu}$, the pion charge form factor is
produced as
\begin{equation}
 {}F_\pi(Q^2) \approx  1 - \frac{ g_{\rho\pi\pi} \; Q^2} 
            {g_\rho \, (Q^2 + m_\rho^2 - im_\rho \, \Gamma_\rho(Q^2))} \;.
 \label{piffvmd}
\end{equation}
The non-resonant first term arises from the photon coupling to the
charge of a point pion.  The resonant second term arises from
\mbox{$\rho-\gamma$} coupling, implements all of the $Q^2$-dependence,
and vanishes at \mbox{$Q^2=0$} in accord with gauge invariance.  The
width $\Gamma_\rho$ is nonzero beyond the threshold for $\pi\,\pi$
production only, and the form factor is real for $Q^2 > -4\,m_\pi^2$.
In this model the charge radius \mbox{$r_\pi^2 = -6 F^\prime_\pi(0)$}
comes entirely from the resonant term and is \mbox{$6
g_{\rho\pi\pi}/(m_\rho^2 g_\rho) = 0.48~{\rm fm}^2$}, which compares
favorably with the experimental value $0.44~{\rm fm}^2$.  See
Ref.~\cite{OPTW95} for a recent review of the pion charge form factor
and VMD in this form.

In terms of QCD, where the pion is a $\bar q q$ bound state, the content
of Eq.~(\ref{piffvmd}) cannot provide a realistic picture for
$F_\pi(Q^2)$ at (low) spacelike $Q^2$.  The photon couples only to the
distributed quark currents in the pion, the vector meson bound state is
not a well-defined concept away from the pole, and the question of a
resonant $\rho$ contribution has to be addressed within the dressed
quark-photon vertex.  This is the topic we explore in this work.  Given
that one can find a convenient (and necessarily model-dependent)
representation of the quark-photon vertex involving direct and resonant
parts, the direct coupling will necessarily produce a distribution
$F^{dir}_\pi(Q^2)$ to replace the first term in Eq.~(\ref{piffvmd}) and
there will be a corresponding contribution to the charge radius.  The
remaining contribution must differ from the second term of
Eq.~(\ref{piffvmd}) because $g_{\rho\pi\pi}$, $g_\rho$, $m_\rho$ and
$\Gamma_\rho$ are well-defined only for the physical on-shell vector
meson bound state.  A measure of the ambiguities involved is provided by
a recent study that modeled the underlying $\bar q q$ substructure of
$\rho$ and $\pi$.  Under the assumption that the $\rho$ Bethe--Salpeter
(BS) amplitude is applicable also at {$Q^2=0$}, it was found that
$g^{\rm eff}_{\rho\pi\pi}(Q^2=0) \approx
g_{\rho\pi\pi}/2$~\cite{PCT98Adl}.  On face value, this suggests that
only about 50\% of $r_\pi^2$ would be attributable to the
$\rho$-resonance mechanism.

In this work we obtain the quark-photon vertex as the solution of the
inhomogeneous Bethe--Salpeter equation (BSE) in ladder truncation with
the dressed quark propagators taken as solutions of the Dyson--Schwinger
equation (DSE) in rainbow truncation.  Such a procedure automatically
incorporates the pole structure of the vertex corresponding to the
vector meson spectrum of the homogeneous BSE.  A previous exploratory
study~\cite{F95} of the coupled DSE-BSE for the quark propagator and the
quark-photon vertex employed a simple infrared dominant form of the
effective gluon propagator in order to utilize closed form expressions
for the dressed quark propagators and the resulting BSE kernel.  That
study was not carried far enough to draw implications for the pion form
factor.

We employ a more realistic model for the effective quark-antiquark
coupling that has recently been shown to reproduce the pion and kaon
masses and decay constants~\cite{MR97} as well as the masses and decay
constants for the vector mesons $\rho$, $\phi$ and K$^\star$ to within
10\%~\cite{MT99}.  There, as well as here, the quark propagators are
consistently dressed through the quark DSE using the same effective
gluon propagator, which ensures that the vector Ward--Takahashi identity
is obeyed and electromagnetic current is conserved.  The model
parameters are all fixed in previous work~\cite{MT99} and constrained
only by $m_\pi$, $m_K$, $f_\pi$ and $\langle\bar q q\rangle$.

Next, we use this quark-photon vertex to investigate the low momentum
behavior of the pion charge form factor.  The results are compared to
those from the Ball--Chiu (BC) Ansatz for the vertex which is commonly
used in such studies.  In comparison with the results from the BSE
solution, use of the BC Ansatz (and variations thereof) underestimates
the charge radius.  The BC Ansatz does not have the vector meson pole
that is naturally present within the BSE solution.  We explore the
extent to which the BSE solution for the vertex can be reproduced by the
addition of a suitable $\rho$-pole term to the BC Ansatz.  We then
assess the performance of this ``resonant-improved'' Ansatz for
the quark-photon vertex.

In Sec.~II we review the formulation that underlies recent studies of
the pion form factor within a modeling of QCD through the DSEs, and
discuss the need for a dressed quark-photon vertex in such models.  The
BSE for the quark-photon vertex is described in Sec.~III and the vector
meson pole contributions are outlined.  Also described there are: the
model used for the effective coupling in ladder truncation, the basis of
covariants used, and the comparison of the numerical solution with the
resonant-improved BC Ansatz.  Our analysis of the low-momentum behavior
of the pion form factor and charge radius is presented in Sec.~IV, and
concluding remarks are given in Sec.~V.

\section{The pion electromagnetic form factor}

\begin{figure}
\centering{\
\epsfig{figure=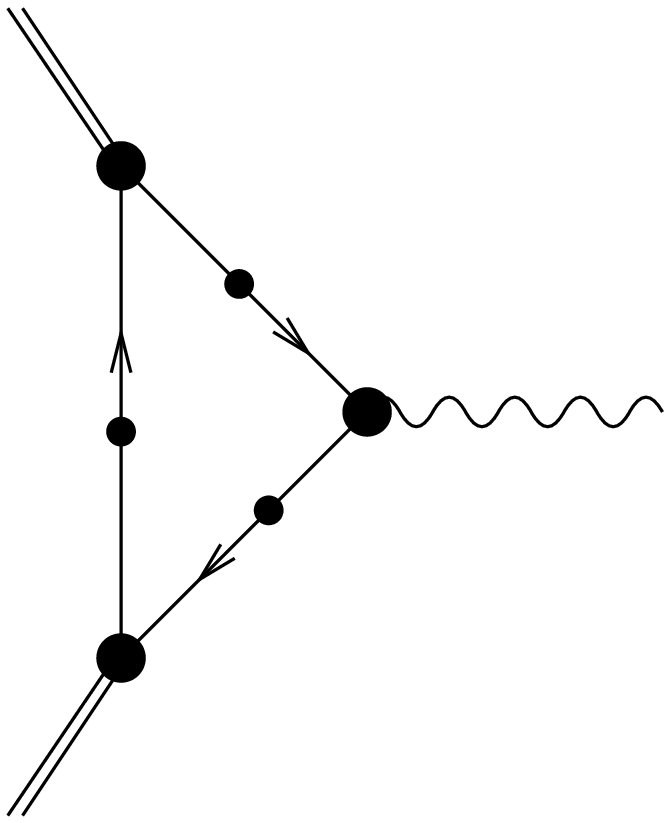,height=6.0cm} }
\caption{ The impulse approximation for the pion charge form factor.
\label{fig:triangle} }
\end{figure}

A number of works have treated the electromagnetic form factors of
pions~\cite{R96,MR98}, kaons~\cite{BRT96,pctrev}, and more recently
vector mesons~\cite{HP99} and nucleons~\cite{BRSBF99} in the course of
QCD modeling via truncations of the DSEs.  Using dressed quark
propagators and bound state BS amplitudes, form factors can be
calculated in impulse approximation, as depicted in
Fig.~\ref{fig:triangle} for the $\gamma\pi\pi$ vertex.  Such work
proceeds most easily in Euclidean metric $\{\gamma_\mu,\gamma_\nu\} =
2\delta_{\mu\nu}$, $\gamma_\mu^\dagger = \gamma_\mu$ and $a\cdot b =
\sum_{i=1}^4 a_i b_i$.  In the space of color, flavor and Dirac spin, we
denote by $\tilde{\Gamma}_\mu(p;Q)$ the quark-photon vertex describing
the coupling of a photon with momentum $Q$ to a quark with initial and
final momenta $p-Q/2$ and $p+Q/2$ respectively.  With this notation, the
$\gamma\pi\pi$ vertex takes the form
\begin{eqnarray}
\label{gpp}
\lefteqn{\Lambda_\nu(P,Q)= 2P_\nu \, F_\pi(Q^2) } 
\nonumber \\ & & 
= -2\, \int\!\frac{d^4q}{(2\pi)^4} \;
       {\rm Tr}\left[ \Gamma_\pi(k_+;-P_+)\; S(q_{+-})\; 
        i\tilde{\Gamma}_\nu(q_+;Q)\; S(q_{++}) \; 
        \Gamma_\pi(k_-;P_-)\; S(q_-) \right] \,,
\end{eqnarray}
where $F_\pi(Q^2)$ is the pion form factor and
\mbox{$P_\pm = P \pm Q/2$}, \mbox{$q_\pm = q \pm P/2$},   
\mbox{$q_{+ \pm} = q_+ \pm Q/2$} and  \mbox{$k_\pm = q \pm Q/4$}.  
$S(q)$ is the dressed quark propagator, $\Gamma_\pi(k;P)$ is the pion BS
amplitude corresponding to relative \mbox{$\bar q q$} momentum $k$ (we
choose equal partitioning) and pion momentum $P$, and $\rm Tr[\ldots]$
denotes the trace over color, flavor and spin indices.  The quark-photon
vertex $\tilde\Gamma_\nu(q;Q)$ at sufficiently large spacelike $Q^2$
becomes
\mbox{$\tilde{\Gamma}_\nu(q;Q) \rightarrow \hat{Q} \, \gamma_\mu$}, 
where $\hat{Q}$ is the quark charge operator.

The quark-photon vertex satisfies the Ward--Takahashi identity (WTI)
\begin{equation}
i\,Q_\mu \,\tilde{\Gamma}_{\mu}(p;Q)  = \hat{Q} 
                       \left( S^{-1}(p+Q/2) - S^{-1}(p-Q/2) \right) \,,
\label{wtid}
\end{equation}
as a result of gauge invariance.  At \mbox{$Q=0$} the vertex is
completely specified by the differential Ward identity
\begin{equation}
i \,\tilde{\Gamma}_{\mu}(p;0) = \hat{Q} \; 
                            \frac{\partial}{\partial p_\mu} S(p)^{-1} \; .
\label{wid}
\end{equation}
This reduces Eq.~(\ref{gpp}) to \mbox{$F_\pi(Q^2=0)=1$} if $\Gamma_\pi$
is properly normalized\footnote{This equivalence holds if the kernel of
the pion BSE is independent of the pion momentum.  For the ladder
truncation of the kernel, which we consider here, this is the case.}.
This clearly shows that the bare vertex $\gamma_\mu$ is generally a bad
approximation due to the momentum-dependence of the quark self-energy:
only with bare quark propagators, does a bare vertex satisfy
Eqs.~(\ref{wtid}) and (\ref{wid}).  Use of a bare vertex, in combination
with dressed propagators, in Eq.~(\ref{gpp}), violates charge
conservation and leads to $F_\pi(0) \neq 1$.

The WTI determines the longitudinal part of the vertex completely in
terms of the (inverse) quark propagator.  However, the transverse part
is largely unconstrained by symmetries.  A common Ansatz for the dressed
vertex is that due to Ball and Chiu~\cite{BC80} which was developed in
the context of QED investigations and is a representation in terms of
the quark propagator functions $A$ and $B$ defined by
\begin{equation}
\label{sinvp}
 S(p)^{-1} = i \,/\!\!\!p A(p^2) + B(p^2)\, .
\end{equation}
With \mbox{$\tilde{\Gamma}^{\rm BC}_\mu(p;Q)=
\hat{Q}\,\Gamma_\mu^{\rm BC}(p;Q)$} the BC Ansatz is
\begin{eqnarray}
 \Gamma_{\mu}^{\rm BC}(p;Q)  &=&
       \case{1}{2}\gamma_\mu\,\Big(A(p_+)+A(p_-)\Big)
        + 2\, /\!\!\!p \, p_\mu\,\frac{A(p_+)-A(p_-)}{p_+^2 - p^2_-}
        - 2\,i\,p_\mu \, \frac{B(p_+)-B(p_-)}{p_+^2 - p^2_-} \; ,
\label{bcver}
\end{eqnarray}
where \mbox{$p_\pm = p \pm Q/2$}.  This satisfies the constraints from
the WTI Eq.~(\ref{wtid}) and the Ward identity Eq.~(\ref{wid}),
transforms under CPT as a vector vertex should, and has the correct
perturbative limit $\gamma_\mu$ in the ultraviolet.  The longitudinal
part of $\Gamma_\mu^{\rm BC}$ is exact; the transverse part is exact
only at \mbox{$Q=0$} and in the UV limit.  Curtis and Pennington have
explored the additional Dirac structures that are possible for the
transverse part of the vertex and have suggested an improved
Ansatz~\cite{CP90} based on multiplicative renormalizibility in QED
\begin{eqnarray}
\lefteqn{ \Gamma_{\mu}^{\rm CP}(p;Q) \;=\; 
        \Gamma_{\mu}^{\rm BC}(p;Q) + {}} 
\nonumber \\ &&
        \Big( (p_+^2 - p_-^2)\,\gamma_\mu 
        - (\,/\!\!\!p_ + - \,/\!\!\!p_-)\,p_\mu \Big)
        \frac{(p_+^2 + p_-^2)(A(p_+)-A(p_-))}
        {(p_+^2-p_-^2)^2 + (M(p_+)^2 + M(p_-)^2)^2}\,,
 \label{cpver}
\end{eqnarray}
with $M(p) = B(p)/A(p)$.  Both Ans\"atze Eq.~(\ref{bcver}) and
Eq.~(\ref{cpver}) for the $\bar q q \gamma$ vertex satisfy all symmetry
constraints, and use of them in Eq.~(\ref{gpp}) leads to $F_\pi(0) = 1$,
but neither of them contain the timelike vector meson poles of the exact
vertex.

The pion charge radius 
\begin{equation}
 r_\pi^2 = -6 \,\frac{\partial F_\pi(Q^2)}{\partial Q^2}\Bigg|_{Q^2=0}\;,
 \label{rpi2}
\end{equation}
will receive two types of contribution within the impulse approximation
in Eq.~(\ref{gpp}): 1) contributions from $\tilde{\Gamma}_\nu(Q=0)$
coupled with the $Q^2$-slope produced by the quark propagators and
$\Gamma_\pi$, and 2) contributions proportional to the $Q^2$-slope of
$\tilde{\Gamma}_\nu(Q)$.  For the first type, use of the BC Ansatz is
sufficient since the BC Ansatz is exact at $Q=0$.  However for the
second type of contribution, \mbox{$\partial\tilde\Gamma_{\mu}(p;Q)/
\partial Q^2 $} is {\em not} constrained by symmetries.

Previous studies of three-point quark loops such as Eq.~(\ref{gpp}) for
form factors have, for practical reasons, utilized parameterized
representations of the DSE solutions for quark propagators and the BS
amplitudes, in conjunction with the BC Ansatz for the quark-photon
vertex~\cite{R96,MR98,BRT96,pctrev,HP99}.  The parameters are fitted to
give a good description of pion and related chiral observables such as
$f_\pi$, $m_\pi$, $r_\pi$, and $\langle\bar q q\rangle$ with a view
towards parameter-free studies of other mesons and
observables~\cite{IKR99}.  This procedure can produce values of
$r^2_\pi$ in the range 20\% to 30\% below the experimental
value~\cite{MR98,IKR99}, which leaves some room for additional
contributions, such as those coming from $\pi\,\pi$
rescattering~\cite{ABR95} and from the $\rho$-resonance mechanism.
However the estimated 50\% of $r_\pi^2$ being due to the
$\rho$-resonance~\cite{PCT98Adl} appears to be incompatible with these
parametrizations.

Here, we reconcile the VMD picture with the QCD picture of a photon
coupled to distributed quarks in a $\bar q q$ bound state by calculating
all ingredients needed for $F_\pi(Q^2)$ via Eq.~(\ref{gpp}) from their
dynamical equations: we solve the quark DSE, the pion BSE, and the
inhomogeneous vertex BSE in a self-consistent way, using the same model
for the effective quark-antiquark coupling.  Both resonance and
non-resonance contributions to the vertex are dynamically generated
through the BSE.  By comparison with the BC Ansatz we can identify
resonance contributions; however, one must keep in mind that such an
identification is necessarily model-dependent.

\section{The Bethe--Salpeter solution for the vector vertex}

The quark-photon vertex satisfies the renormalized inhomogeneous BSE 
\begin{equation}
\tilde{\Gamma}_\mu(p;Q) = Z_2\; \hat{Q}\; \gamma_\mu + 
        \int^\Lambda\!\!\frac{d^4q}{(2\pi)^4} \; K(p,q;Q) 
         S(q+\eta Q)\tilde{\Gamma}_\mu(q;Q) S(q-\bar\eta Q) \;,
\label{verBSE}
\end{equation}
where \mbox{$\eta + \bar\eta = 1$} describes the momentum sharing
between the two quarks.  The kernel $K$ operates in the direct product
space of color, flavor and Dirac spin for the quark and antiquark and is
the renormalized, amputated $\bar q q$ scattering kernel that is
irreducible with respect to a pair of $\bar q q$ lines.  The notation
$\int^\Lambda$ denotes a translationally-invariant regularization of the
integral, with $\Lambda$ the regularization mass-scale.  At the end of
all calculations the regularization is removed by taking the limit
$\Lambda \to \infty$.

The renormalization constant $Z_2$ and the renormalized dressed quark
propagator $S$ follow from the quark DSE
\begin{eqnarray}
\label{gendse}
 S(p)^{-1} & = & Z_2\,i\,/\!\!\!p + Z_4\,m(\mu)
        + Z_1 \int^\Lambda \frac{d^4q}{(2\pi)^4} \,g^2 D_{\mu\nu}(p-q) 
        \frac{\lambda^a}{2}\gamma_\mu S(q)\Gamma^a_\nu(q,p) \,,
\end{eqnarray}
where $D_{\mu\nu}(k)$ is the renormalized dressed-gluon propagator and
$\Gamma^a_\nu(q;p)$ is the renormalized dressed-quark-gluon vertex.
The solution of Eq.~(\ref{gendse}) is renormalized according to
\begin{equation}
\label{renormS}
S(p)^{-1}\bigg|_{p^2=\mu^2} = i\,/\!\!\!p + m(\mu)\,,
\end{equation}
at a sufficiently large spacelike $\mu^2$, with $m(\mu)$ the
renormalized quark mass at the scale $\mu$.  In Eq.~(\ref{gendse}), $S$,
$\Gamma^a_\mu$ and $m(\mu)$ depend on the quark flavor, although we have
not indicated this explicitly.  The renormalization constants depend on
the renormalization point and the regularization mass-scale, but not on
flavor: in our analysis we employ a flavor-independent renormalization
scheme.

\subsection{Bound state contributions}

Solutions of the homogeneous version of Eq.~(\ref{verBSE}) at discrete
timelike momenta $Q^2$ define vector meson bound states with masses
\mbox{$m_n^2=-Q^2$}.  It follows that $\tilde{\Gamma}_\mu(p;Q)$ has
poles at those locations.  The corresponding resonant form can be
obtained by observing that Eq.~(\ref{verBSE}) has the equivalent form
\begin{equation}
\tilde{\Gamma}_\mu(p;Q) = Z_2\; \hat{Q}\;\gamma_\mu + 
        Z_2 \, \int^\Lambda\!\!\frac{d^4q}{(2\pi)^4} \; M(p,q;Q) 
         S(q+\eta Q)\; \hat{Q}\gamma_\mu \; S(q-\bar\eta Q) \;,
\label{verM}
\end{equation}
with $M$ being the $\bar q q$ scattering amplitude given in schematic
form by \mbox{$M=K + K SMS$}.  In the vicinity of the timelike points
\mbox{$Q^2 = -m_n^2$} where the homogeneous BSE has solutions, 
$M$ has poles with residues that define the physical meson BS
amplitudes.  In particular, for two flavors,
\begin{equation}
 M(p,q;Q) \rightarrow - \frac{\Gamma_\nu^\rho(p;Q) 
                \bar{\Gamma}_\nu^\rho(q;-Q)}{Q^2 + m_\rho^2} 
             - \frac{\Gamma_\nu^\omega(p;Q) \bar{\Gamma}_\nu^\omega(q;-Q)}
                              {Q^2 + m_\omega^2} \; , 
\label{Mres}
\end{equation}  
with the BS amplitudes properly normalized.  In the ladder truncation
that we will be concerned with in practice, the normalization condition
reduces to (\mbox{$n = \rho^0, \omega$})
\begin{eqnarray}
\label{vecnorm}
    \left. 2 P_\mu = \frac{\partial}{\partial P_\mu} \,
        \frac{1}{3} \int^\Lambda\!\!\frac{d^4q}{(2\pi)^4}
        {\rm Tr}\left[ \bar\Gamma_\nu^n(q;-K)\, S(q+\eta P)\, 
        \Gamma_\nu^n(q;K)\,S(q-\bar\eta P) \right]
        \right|_{P^2=K^2=-m^2} \,,
\end{eqnarray}
where the factor $1/3$ appears because the three transverse directions
are summed.  Using the fact that massive vector mesons are transverse,
the resonant form of the quark-photon vertex near the vector meson poles
can be written as
\begin{equation}
\tilde{\Gamma}_\mu(p;Q) \rightarrow   \frac{\Gamma_\mu^\rho(p;Q) 
                          m_\rho^2/g_\rho }{Q^2 + M_\rho^2}
                            +  \frac{\Gamma_\mu^\omega(p;Q) 
                          m_\omega^2/g_\omega }{Q^2 + M_\omega^2} \; ,
\label{verres}
\end{equation}
where the coupling constants $g_n$ (\mbox{$n=\rho^0, \omega$}) for
$\rho-\gamma$ and $\omega-\gamma$ mixing are
\begin{equation}
\label{rhophoton}
   \frac{m_n^2}{g_n} = - \frac{Z_2}{3}
        \int^\Lambda\!\!\frac{d^4q}{(2\pi)^4} \;
        {\rm Tr}\left[\bar{\Gamma}^n_\nu(q;-Q) 
        S(q+\eta Q) \; \hat{Q} \gamma_\nu \; S(q-\bar\eta Q)\right] \; .
\end{equation}
Away from the pole, the separation into a resonant and non-resonant part is 
not unique; the solution of the BSE for the complete vertex contains both 
aspects in a consistent way. 

At the level of the ladder approximation, which is commonly used in
practical calculations, and using isospin symmetry, we have
\mbox{$m_\rho = m_\omega = m_V$}, and the flavor structure of the BS
amplitudes is \mbox{$\Gamma_\nu^\rho(p;Q)= (\tau_3/\sqrt{2})
\Gamma_\nu^V(p;Q)$}, and \mbox{$\Gamma_\nu^\omega(p;Q)= (1/\sqrt{2})
\Gamma_\nu^V(p;Q)$}.  The flavor trace in Eq.~(\ref{rhophoton}) gives
\mbox{$g_\omega = 3 g_\rho = 3 g_V$}.  The factorization 
\mbox{$\tilde{\Gamma}_\mu = \hat{Q} \Gamma_\mu$} allows Eq.~(\ref{verres})
for the vector pole structure to simplify to the flavor-independent form
\begin{equation}
 \Gamma_\mu(p;Q) \rightarrow 
        \frac{\Gamma_\mu^V(p;Q) f_V m_V}{Q^2 + m_V^2} \; ,
\label{comres}
\end{equation}
where \mbox{$f_V m_V = \sqrt{2} m_V^2 /g_V $}.  Our numerical study is
carried out for the flavor-independent vector vertex $\Gamma_\mu(p;Q)$.

\subsection{Ladder--rainbow truncation}

We use a ladder truncation for the BSE 
\begin{equation}
\label{ourBSEansatz}
        K^{rs}_{tu}(p,q;P) \to
        -{\cal G}((p-q)^2)\, D_{\mu\nu}^{\rm free}(p-q)
        \left(\frac{\lambda^a}{2}\gamma_\mu\right)^{ru} \otimes
        \left(\frac{\lambda^a}{2}\gamma_\nu\right)^{ts} \,,
\end{equation}
where $D_{\mu\nu}^{\rm free}(k)$ is the perturbative gluon propagator in
Landau gauge. The resulting BSE is consistent with a rainbow truncation
\mbox{$\Gamma^a_\nu(q,p) \rightarrow \gamma_\nu \lambda^a/2$} for the 
quark DSE, Eq.~(\ref{gendse}), in the sense that the combination
produces vector and axial-vector vertices satisfying the respective
WTIs.  In the axial case, this ensures that in the chiral limit the
ground state pseudoscalar mesons are massless even though the quark mass
functions are strongly enhanced in the infrared~\cite{MR97,MRT98}.  In
the vector case, this ensures electromagnetic current conservation.

The model is completely specified once a form is chosen for the
``effective coupling'' ${\cal G}(k^2)$.  We employ the
Ansatz~\cite{MR97,MT99}
\begin{equation}
\label{gvk2}
\frac{{\cal G}(k^2)}{k^2} =
               \frac{4\pi^2}{\omega^6} D k^2 {\rm e}^{-k^2/\omega^2}
+ 4\pi\,\frac{ \gamma_m \pi}    {\case{1}{2}
        \ln\left[\tau + \left(1 + k^2/\Lambda_{\rm QCD}^2\right)^2\right]}
{\cal F}(k^2) \,,
\end{equation}
with ${\cal F}(k^2)= [1 - \exp(-k^2/[4 m_t^2])]/k^2$, and
\mbox{$\gamma_m=12/(33-2N_f)$}.  This Ansatz preserves the one-loop
renormalization group behavior of QCD for solutions of the quark DSE.
In particular, it produces the correct one-loop QCD anomalous dimension
of the quark mass function $M(p^2)$ for both the chiral limit and
explicit chirally broken case~\cite{MR97,asbeh}.  The first term of
Eq.~(\ref{gvk2}) implements the strong infrared enhancement in the
region \mbox{$k^2=0-1$}~GeV$^2$ which is a phenomenological requirement
for sufficient dynamical chiral symmetry breaking to produce an
acceptable strength for the quark condensate~\cite{HMR98}.  We use
\mbox{$m_t=0.5$~GeV},
\mbox{$\tau={\rm e}^2-1$}, \mbox{$N_f=4$},
\mbox{$\Lambda_{\rm QCD}^{N_f=4}= 0.234\,{\rm GeV}$}, and a
renormalization point $\mu=19\,$GeV, which is sufficiently perturbative
to allow the one-loop asymptotic behavior of the quark propagator to be
used as a check~\cite{MR97,MT99}.  The remaining parameters are fixed to
\mbox{$\omega = 0.4~{\rm GeV}$} and
\mbox{$D=0.93~{\rm GeV}^2$} to give a good description of $m_{\pi/K}$
and $f_{\pi}$.  The subsequent values for $f_K$ and the masses and decay
constants of the vector mesons $\rho, \phi, K^\star $ are very well
described~\cite{MT99}.

\subsection{Numerical solution}

The general form of $\Gamma_\mu(q;Q)$ can be decomposed into twelve
independent Lorentz covariants, made from the three vectors
$\gamma_\mu$, the relative momentum $q_\mu$, and the photon momentum
$Q_\mu$, each multiplied by one of the four independent matrices
$1\hspace{-3pt}$l, $\,/\!\!\!q$, $\;/\!\!\!\!Q$, and $\sigma_{\mu\nu}
q_\mu Q_\nu$.  Four of the covariants represent the longitudinal
components which are completely specified by the WTI and can be taken as
the longitudinal projection $\Gamma_\mu^L$ of the BC Ansatz,
Eq.~(\ref{bcver}).  The solution of the BSE for the transverse vertex
can be expanded in eight covariants $T^i_\mu(q;Q)$.  Thus the total
vertex is decomposed as
\begin{equation}
\label{genvecbsa}
 \Gamma_\mu(q;Q)  =  \Gamma_\mu^L(q;Q) + \sum_{i=1}^8 \, T^i_\mu(q;Q) \, 
                     F_i(q^2, q\cdot Q; Q^2)\,,
\label{verdecom}
\end{equation}
with the invariant amplitudes $F_i$ being Lorentz scalar functions.  The
choice for the covariants $T^i_\mu(q;Q)$ to be used as a basis is
constrained by the required properties under Lorentz and CPT
transformations, but is not unique.  The BSE, Eq.~(\ref{verBSE}), must
be projected onto the covariant basis to produce a coupled set of eight
linear equations for the invariant amplitudes $F_i$ to be cast in matrix
form.  This requires a procedure to project out a single amplitude from
the general form Eq.~(\ref{genvecbsa}).  It is therefore helpful if the
chosen covariants satisfy a Dirac-trace orthogonality property. The
following set of orthogonal covariants is used here\footnote{Since the
domain of interest here includes both timelike and spacelike $Q^2$, we
avoid factors of $Q = \sqrt{Q^2}$.  Therefore we do not require the
covariants to be normalized, as we did in Ref.~\cite{MT99}.  We also
include explicit factors of $q\cdot Q$ in $T_3$ and $T_6$ so that now
every $F_i$ is even in $q\cdot Q$.}
\begin{eqnarray}
        T^1_\mu(q;Q) & = & \gamma^T_\mu                         
\label{cov1}    \,,\\
        T^2_\mu(q;Q) & = & 
                \left( q^T_\mu \;/\!\!\!q^T - 
                \case{1}{3} \gamma^T_\mu (q^T)^2 \right)/q^2      \,,\\
        T^3_\mu(q;Q) & = & 
                 q^T_\mu \;/\!\!\!\!Q \, q \cdot Q/(q^2 Q^2)      \,,\\
        T^4_\mu(q;Q) & = &         
               - \left(\gamma^T_\mu [\;/\!\!\!\!Q,\,/\!\!\!q] 
                + 2\,q^T_\mu \;/\!\!\!\!Q \right)/2q           
\label{cov4}   \,,\\   
        T^5_\mu(q;Q) & = & i\; q^T_\mu/q                          \,,\\
        T^6_\mu(q;Q) & = & i\;
                [\gamma_\mu^T , \;/\!\!\!q^T ] \, q \cdot Q/q^2   \,,\\ 
        T^7_\mu(q;Q) & = & i\,
        [\gamma_\mu^T , \;/\!\!\!\!Q] \left(1 - \cos^2{\theta}\right)
                                      - 2\;T^8_\mu(q;Q)           \,,\\
        T^8_\mu(q;Q) & = & i\;
                q^T_\mu \;/\!\!\!q^T\,\;/\!\!\!\!Q/q^2        \,,
\label{cov8}
\end{eqnarray}
where $V^T$ is the component of $V$ transverse to $Q$, that is
\mbox{$V^T_\mu =$} \mbox{$V_\mu - Q_\mu\,(Q\cdot V)/Q^2$}
and $q\cdot Q = q\,Q\,\cos\theta$.  Each amplitude can be projected onto
a basis of even-order Chebyshev polynomials in $\cos\theta$ to produce
$^n F_i(q^2, Q^2)$~\cite{MT99}.  It is found that terms beyond the
zeroth order, $n > 0$, are insignificant and are not used in the
subsequent study of the form factor.

\begin{figure}
\centering{\
\epsfig{figure=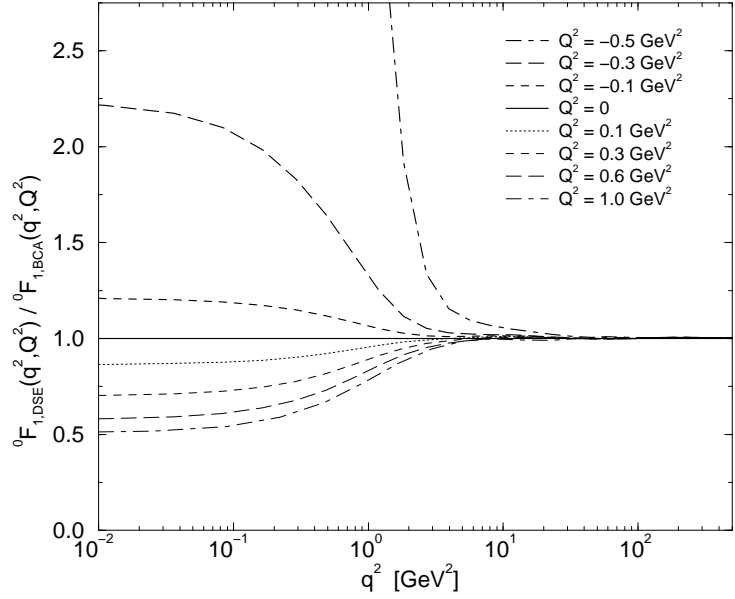,height=8.0cm} }
\caption{
The ratio of the BSE solution over the BC Ansatz for the dominant
amplitude ${}^0F_1$, associated with $\gamma^T_\mu$, of the quark-photon
vertex as function of $q^2$ for the indicated values of the photon
momentum $Q^2$.  The quark momenta are $q\pm Q/2$.
\label{fig:F1q} }
\end{figure}

For a range of small timelike and spacelike $Q^2$, the ratio of the BSE
solution to the BC Ansatz for the dominant amplitude ${}^0F_1$ is
displayed in Fig.~\ref{fig:F1q} as function of $q^2$.  The deviation
from unity indicates differences between the BC Ansatz and our numerical
solution of the ladder BSE.  It is immediately obvious that for timelike
photon momenta and \mbox{$q^2 < 3~{\rm GeV}^2$} there is a significant
enhancement of the BSE solution compared to the BC Ansatz.  This strong
increase in the vertex BSE solution is due to the vector meson pole at
\mbox{$Q^2 \approx -0.55~{\rm GeV}^2$}, which only affects the vertex at
small $q^2$; at large $q^2$ both the BSE solution and the BC Ansatz
approach the bare vertex for any value of $Q^2$.

Charge conservation is evident through the ratio being one at $Q=0$: the
ladder vertex BSE solution satisfies the WTI.  The fact that our
numerical solution indeed is to within 1\% equal to the BC Ansatz, not
only for ${}^0F_1$, but also for the other amplitudes, indicates the
accuracy of our numerical methods.  At small spacelike momenta the BSE
solution differs considerably from the BC Ansatz.  The difference
increases with $Q^2$ in the range $0 < Q^2 < 2~{\rm GeV}^2$; it is only
at asymptotic values of $Q^2$ that they become equal again.

\begin{figure}
\centering{\
\epsfig{figure=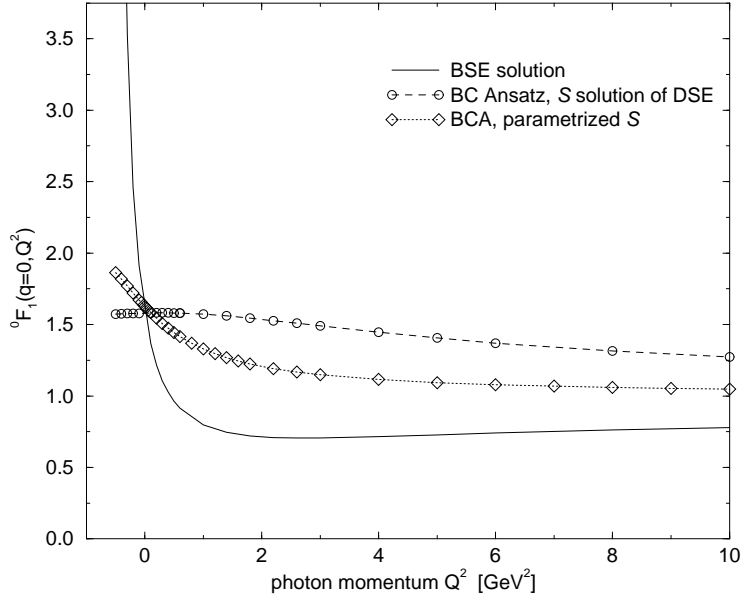,height=8.0cm} }
\caption{  
The $Q^2$-dependence of the BSE solution for the dominant amplitude
${}^0F_1$ of the $\bar q q\gamma$ vertex at \mbox{$q^2=0$} (solid line).
The dashed line with circles is the BC Ansatz for this amplitude in the
present model; the dotted line with diamonds is the BC Ansatz with
phenomenological quark propagator functions $A$ and $B$ parameterized
and fitted to pion observables, including $r_\pi$, according to
Ref.~\protect\cite{MR98}.
\label{fig:F1Q} }
\end{figure}

A different perspective is shown by Fig.~\ref{fig:F1Q} which compares
the $Q^2$-dependence of ${}^0F_1$ at \mbox{$q^2=0$}.  Again, the
agreement at \mbox{$Q^2=0$} is dictated by the Ward identity; however,
the slope of ${}^0F_1(0,Q^2)$ at $Q^2=0$ as obtained from the vertex BSE
solution differs significantly from that produced by the BC Ansatz.
This can be of importance for the charge radii of the pion and other
hadrons.  In this figure one can clearly see the presence of the vector
pole in the $\bar q q \gamma$ vertex at \mbox{$Q^2 \approx -0.55~{\rm
GeV}^2$}.  Also in this figure we plot the amplitude ${}^0F_1$ as
obtained from the BC Ansatz with the quark propagators of
Ref.~\cite{MR98}, which were parameterized rather than obtained as the
solution of a DSE.  The parameters were fitted to $f_\pi,$ $m_\pi$,
$r_\pi$, and $\langle\bar q q\rangle$, using the BC Ansatz in the
calculation of $r_\pi$.  The $Q^2$-dependence of the $\bar q q \gamma$
vertex necessary in order to reproduce a reasonable value for $r_\pi$
was thus incorporated via the quark propagator functions.  We see that
this phenomenological vertex amplitude has indeed a much stronger
$Q^2$-dependence than is obtained by use of the present DSE solution for
the quark propagator in the BC Ansatz.  This phenomenological procedure
allows one to get reasonable electromagnetic form factors at low $Q^2$
without explicitly taking resonances into account.

We have investigated the extent to which the vertex BSE solution can be
represented by the interpolating form
\begin{eqnarray}
\label{bcplus}
   \Gamma_\mu(q;Q) &\simeq& \Gamma_\mu^{\rm BC}(q;Q) -
        \sum_i \, T^i_\mu(q;Q) \, F_i^V(q^2)\;
              \frac{f_V \; Q^2}{m_V(Q^2 + m_V^2)}\, \; ,
\end{eqnarray}
where $F_i^V$ are the leading Chebyshev moments of the mass-shell vector
meson BS amplitudes produced by the present model~\cite{MT99}.  The
second term Eq.~(\ref{bcplus}) correctly describes the vertex near the
vector meson pole at \mbox{$Q^2= -m_V^2$}; there is no width generated
for the vector meson in the present ladder truncation of the BSE.  One
would have to add the $\pi\pi$ channel to the ladder BSE kernel to
produce a $\rho$ width; for the vertex, this would generate an imaginary
part beyond the threshold for pion production, $Q^2 < -4m_\pi^2$.  

\begin{figure}
\centering{\
\epsfig{figure=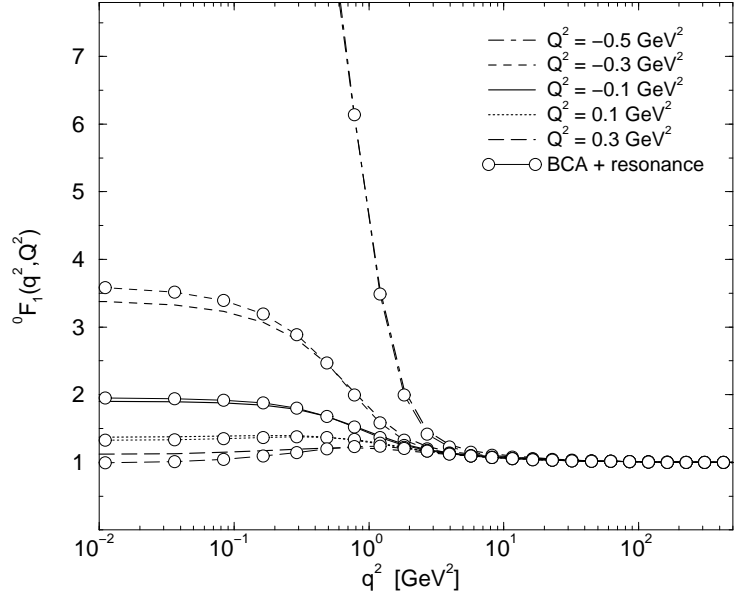,height=8.0cm} }
\caption{ 
The $q^2$-dependence of the BSE solution for the dominant amplitude
${}^0F_1$ for a range of timelike and spacelike values of the photon
momentum $Q^2$.  The circles correspond to the resonant-improved BC
Ansatz as described in the text.
\label{fig:F1BCres} }
\end{figure}

\begin{figure}
\centering{\
\epsfig{figure=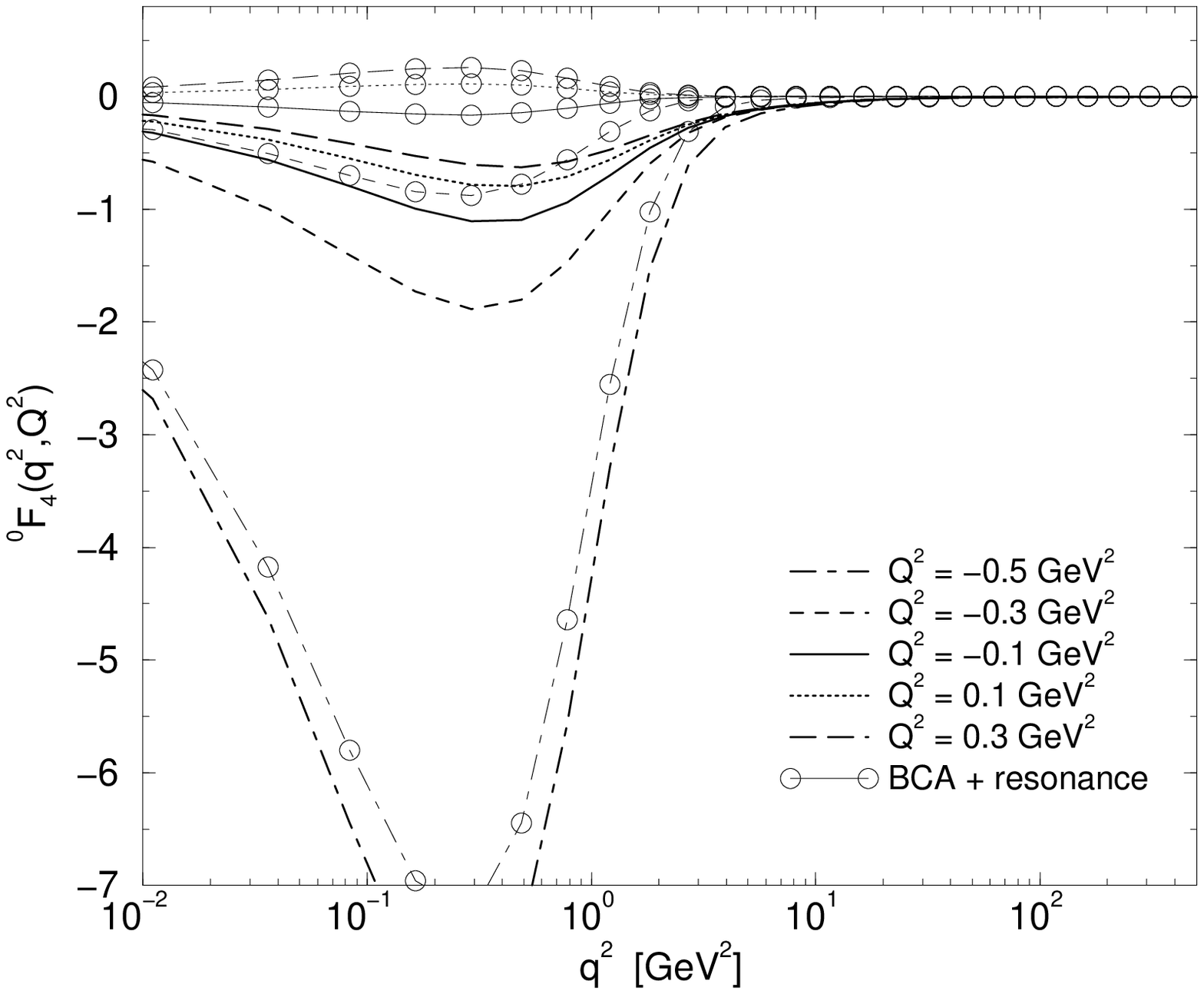,height=8.0cm} }
\caption{ 
The same as Fig.~\protect\ref{fig:F1BCres} except now for the
$\bar q q\gamma$ vertex amplitude ${}^0F_4$.
\label{fig:F4BCres} }
\end{figure}

\begin{figure}
\centering{\
\epsfig{figure=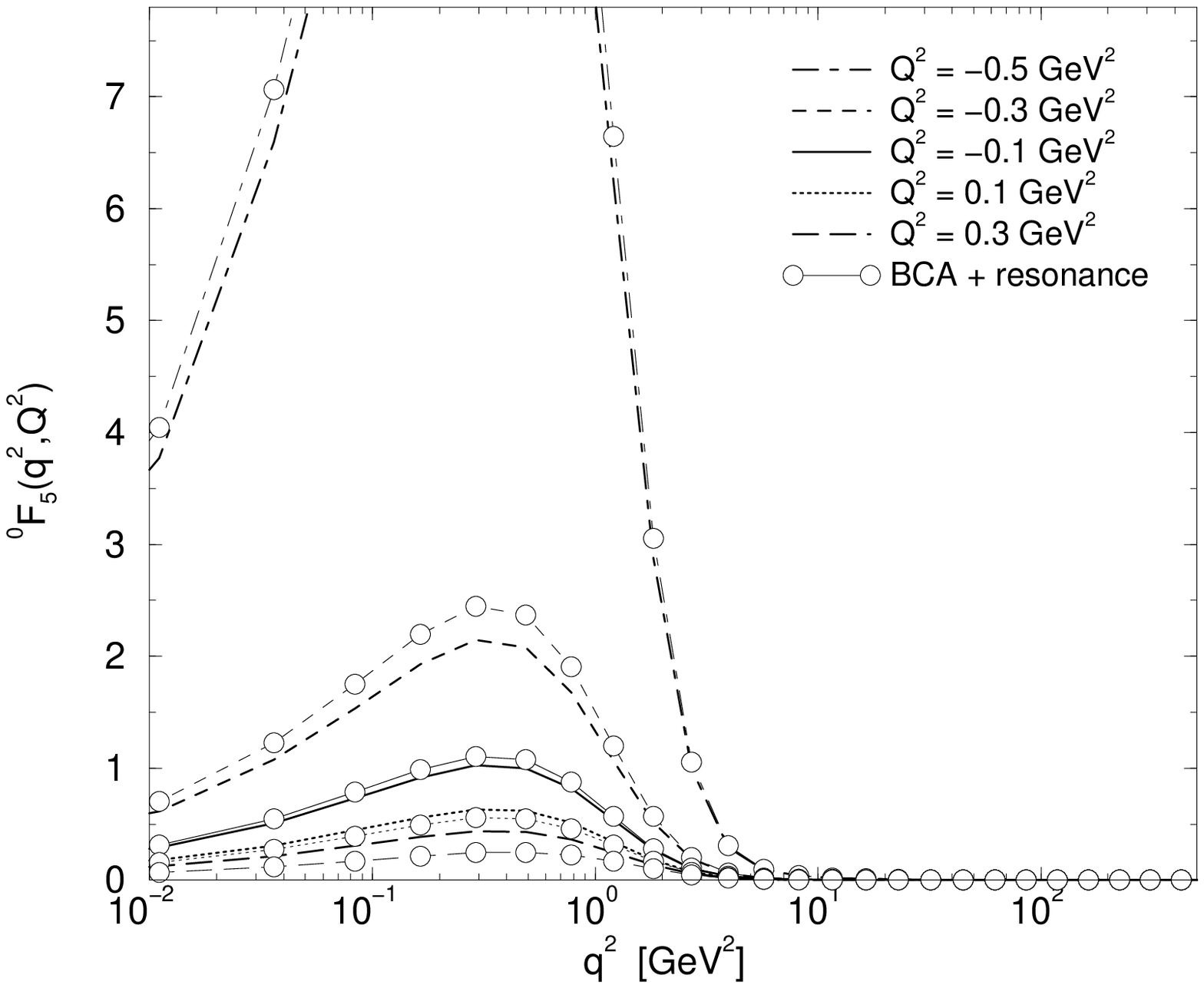,height=8.0cm} }
\caption{
The same as Fig.~\protect\ref{fig:F1BCres} except now for the
$\bar q q\gamma$ vertex amplitude ${}^0F_5$.
\label{fig:F5BCres} }
\end{figure}

For small spacelike $Q^2$, Eq.~(\ref{bcplus}) represents an
extrapolation consistent with: a) $\Gamma^{\rm BC}(q;Q=0)$ is exact due
to gauge invariance and b) the $\rho-\gamma$ coupling should not
generate a photon mass.  Over the limited domain \mbox{$-m_V^2 < Q^2 <
0.2$}~GeV$^2$, Eq.~(\ref{bcplus}) produces a very good representation of
the vertex amplitudes $^0F_1$, $^0F_3$ and $^0F_5$ and a quite
reasonable representation for the remainder, except for $^0F_4$ which is
clearly the worst case, and the hardest to represent in a resonance
formula.  In Figs.~\ref{fig:F1BCres}, \ref{fig:F4BCres} and
\ref{fig:F5BCres}, this approximate form is compared to the vertex BSE
solution for the three dominant amplitudes $^0F_1$, $^0F_4$ and $^0F_5$.

In this sense we shall refer to the BC Ansatz as missing a contribution
near the $\rho$ pole, and to Eq.~(\ref{bcplus}) as a resonant-improved
BC Ansatz applicable in the region \mbox{$-0.4~{\rm GeV}^2 < Q^2 <
0.2~{\rm GeV}^2$} only.  At asymptotic $Q^2$, this form is obviously not
adequate: it does not go to the bare vertex.  Close to the $\rho$ pole,
neither our vertex BSE solution nor the interpolating form
Eq.~(\ref{bcplus}) are physically realistic, because neither one takes
into account the width $\Gamma_\rho$ generated by the open $\pi\,\pi$
decay channel.  Since $\Gamma_\rho/m_\rho \sim 0.2$, the width
contributes little for $Q^2 > -0.4~{\rm GeV}^2$, and the $\rho$
contribution can be approximated by the simple pole term in
Eq.~(\ref{bcplus}) for applications such as the pion form factor
calculation.

\section{Results for the pion form factor}

In addition to the solution of the BSE for the $\bar q q
\gamma$ vertex and the quark propagator $S$ as the solution of its DSE,
we also need the pion BS amplitude $\Gamma_\pi(q;P)$.  This BS amplitude
is the solution of a homogeneous BSE, and for pseudoscalar bound states
it has the form~\cite{MR97}
\begin{equation}
\label{genpion}
 \Gamma_{\pi}(q;P)  =  \gamma_5 \left[ i E_\pi(q;P) + 
        \;/\!\!\!\! P \, F_\pi(q;P) + 
        \,/\!\!\!q \,q\cdot P\, G_\pi(q;P) +
        \sigma_{\mu\nu}\,q_\mu P_\nu \,H_\pi(q;P) \right]\,,
\end{equation}
with the invariant amplitudes $E_\pi$, $F_\pi$, $G_\pi$ and $H_\pi$
being Lorentz scalar functions of $q^2$ and $q\cdot P=q P\,\cos\theta$,
with $P^2= -m_\pi^2$ the fixed, on-shell, pion momentum.  Each amplitude
can be projected onto a basis of even-order Chebyshev polynomials in
$\cos\theta$ and in Ref.~\cite{MR97} it was found that terms beyond the
zeroth order are insignificant.  With these ingredients we can now
calculate the pion electromagnetic form factor in impulse approximation,
Eq.~(\ref{gpp}).

In Fig.~\ref{fig:piFF} we show our results for the pion form factor for
a range of timelike and spacelike $Q^2$ using the BSE solution for the
$\bar q q \gamma$ vertex, and compare our results with various other
treatments of this vertex.  Clearly a bare quark-photon vertex is
incorrect: current conservation, which ensures \mbox{$F_\pi(0)=1$}, is
violated.  Use of the BC Ansatz conserves the current because it
satisfies the vector Ward identity.  However the resulting form factor
misses the data completely; the $Q^2$-dependence is clearly too small.
On the other hand, the BSE solution for the vertex gives an excellent
description of the low $Q^2$ data, both in the spacelike and timelike
region, {\em without fine tuning the model parameters}: the parameters
are completely fixed by $m_{\pi}$ and $f_\pi$ in Ref.~\cite{MT99}.  The
use of the Curtis--Pennington Ansatz, Eq.~(\ref{cpver}), gives results
for the form factor essentially the same as those obtained with the BC
Ansatz.

Experimentally, the form factor shows a resonance peak at $Q^2 =
-m_\rho^2 = -0.593~{\rm GeV}^2$.  Our calculated $F_\pi(Q^2)$ using the
BSE solution diverges as $1/(Q^2 + m_V^2)$ as one approaches $Q^2 =
-m_V^2 \approx -0.55~{\rm GeV}^2$, where the homogeneous BSE admits a
bound state solution: in the ladder truncation for the BSE one does not
get a width for the $\rho$ meson.  We expect that if we incorporate the
open decay channel $\rho \rightarrow \pi\pi$ in the BSE kernel to
generate a width $\Gamma_\rho$ for the $\rho$, we will have better
agreement with the data close to the resonance peak.  In addition, some
of the present difference between our calculation and the data in the
resonance region is due to the fact that our numerical value for
$m_\rho$ is about 4\% below the physical $\rho$ mass.

Four of the eight amplitudes, $F_3$, $F_6$, $F_7$, and $F_8$, contribute
less than 1\% to $F_\pi(Q^2)$ on the $Q^2$-range considered.  Although
the amplitude $F_3$ contributes very little to the form factor, this
amplitude is needed for strict charge conservation.  Therefore we can
truncate the BSE solution to the dominant five covariants $F_1$ through
$F_5$ without significantly changing the description of $F_\pi(Q^2)$.
In Fig.~\ref{fig:piFF}, the result using the five dominant amplitudes is
almost indistinguishable from the results obtained with all eight
amplitudes.

\begin{figure}
\centering{\
\epsfig{figure=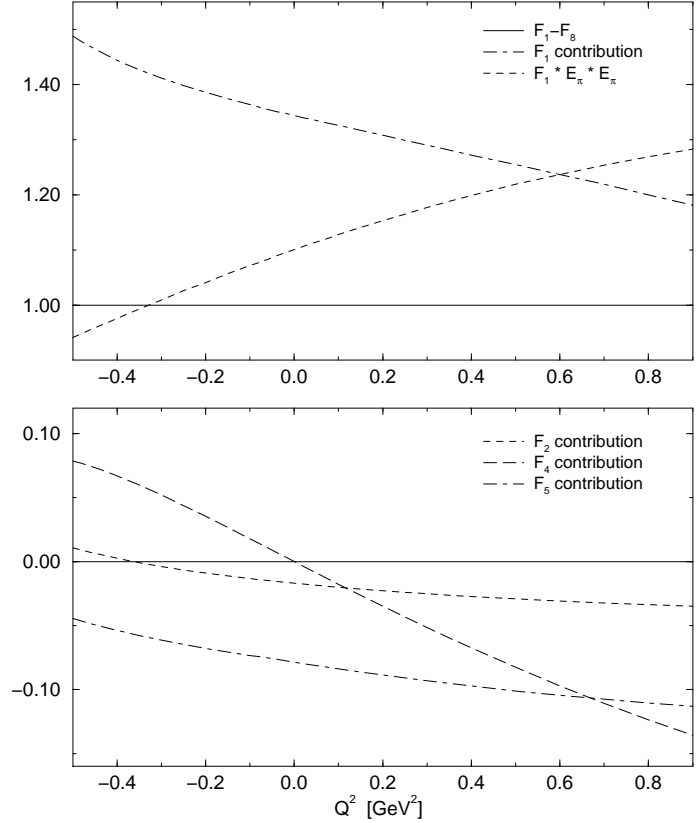,height=11.0cm} }
\caption{ 
The relative contributions to $F_\pi(Q^2)$ from the four most important
amplitudes $F_1$ (top), $F_2$, $F_4$, and $F_5$ (bottom) of the BSE
solution for the $\bar q q\gamma$ vertex.  For the dominant amplitude
$F_1$, we also display the relative contribution from the dominant pion
amplitudes $E_\pi$ only (top).
\label{fig:piFFrel} }
\end{figure}

The relative contributions to $F_\pi(Q^2)$ from the amplitudes $F_1$,
$F_2$, $F_4$, and $F_5$ of the vertex BSE solution are displayed in
Fig.~\ref{fig:piFFrel}.  This shows that the canonical amplitude $F_1$,
together with the canonical (pseudoscalar) amplitude $E_\pi$ of the
pion, generates the bulk of the form factor in the infrared region.  The
pion pseudovector amplitudes $F_\pi$ and $G_\pi$ give a negative
contribution of about 25\% near $Q^2=0$, but they are known to dominate
at large $Q^2$: the asymptotic behavior of the form factor, $F_\pi(Q^2)
\sim 1/Q^2$, is governed by the pseudovector amplitudes of the
pion~\cite{MR98}.  The vertex amplitudes $F_2$, $F_4$, and $F_5$ give
negative contributions to $F_\pi$ of the order of 10\% at small $Q^2$.
Of particular note is the contribution from amplitude $F_4$.  Due to the
Ward identity, this vertex contribution, and hence its contribution to
$F_\pi(Q^2)$, must vanish at \mbox{$Q^2=0$}.  The latter is indeed
evident in Fig.~\ref{fig:piFFrel}; it is also evident there that this
amplitude nevertheless is one of the important contributors to the slope
and hence $r_\pi$; its contribution to the form factor grows roughly
linear with $Q^2$ in the range we studied.  This could have significant
consequences for the form factor at intermediate energies.  At
asymptotic values of $Q^2$ the $\bar q q \gamma$ vertex approaches the
bare vertex, and thus all functions $F_i$ except for $F_1$ will vanish.
The asymptotic behavior of the form factor is therefore not influenced
by $F_4$ nor any other subdominant amplitude of the $\bar q q \gamma$
vertex.

In terms of the pion charge radius $r_\pi$, these results are summarized
in Table~\ref{piradii}.  It is seen that, compared to the empirical
value \mbox{$r_\pi^2=0.44~{\rm fm}^2$}, the BSE solution for the vertex
generates an excellent value: \mbox{$r_\pi^2=0.46~{\rm fm}^2$} using all
eight transverse covariants or using the dominant five, $F_1$ through
$F_5$.  The BC Ansatz produces a value for $r_\pi^2$ that is less than
half this value; this is also evident in Fig.~\ref{fig:piFF}.  The main
reason for this is that the $\bar q q \gamma$ vertex at small but
non-zero $Q^2$ is poorly represented by the BC Ansatz, as shown, for
example, in Fig.~\ref{fig:F1Q}.  The Curtis--Pennington Ansatz gives
almost the same charge radius as the BC Ansatz.

\subsection{Resonance contribution to $r_\pi$}

In the previous section we found that the vertex BSE solution could be
simulated by a resonant-improved BC Ansatz, Eq.~(\ref{bcplus}).  The use
of such a form for the $\bar q q \gamma$ vertex yields, via the impulse
approximation Eq.~(\ref{gpp}), a charge form factor that can be
expressed as
\begin{equation}
 {}F_\pi(Q^2) \approx F^{\rm BC}_\pi(Q^2) - 
         \frac{ g_{\rho\pi\pi} \; F_{V \pi\pi}(Q^2) \; Q^2} 
         {g_\rho (Q^2 + m_\rho^2) } \,,
\label{piffbc+}
\end{equation}
where $F^{\rm BC}_\pi(Q^2)$ is the result from the BC Ansatz. The
combination \mbox{$g_{\rho\pi\pi} F_{V \pi\pi}$} represents the $\pi\pi$
coupling to the vector $\bar q q$ correlation $T_\mu^i F^V_i$ evident in
the resonant term of Eq.~(\ref{bcplus}).  The behavior of this latter
term guarantees that at the vector meson pole, the second term of
Eq.~(\ref{piffbc+}) reproduces the standard behavior and
\mbox{$F_{V \pi\pi}(-m_\rho^2)=1$}.  For other $Q^2$ values $F_{V
\pi\pi}(Q^2)$ does not correspond to a physical process: off-shell
mesons are by definition not physical.  However, the departure of
$F_{V\pi\pi}(Q^2)$ from unity is a rough measure of the difference in
$\pi\pi$ coupling experienced by the effective vector $\bar q q$
correlation away from the $\rho$ mass-shell compared to the physical
$\rho\pi\pi$ coupling.

\begin{figure}
\centering{\
\epsfig{figure=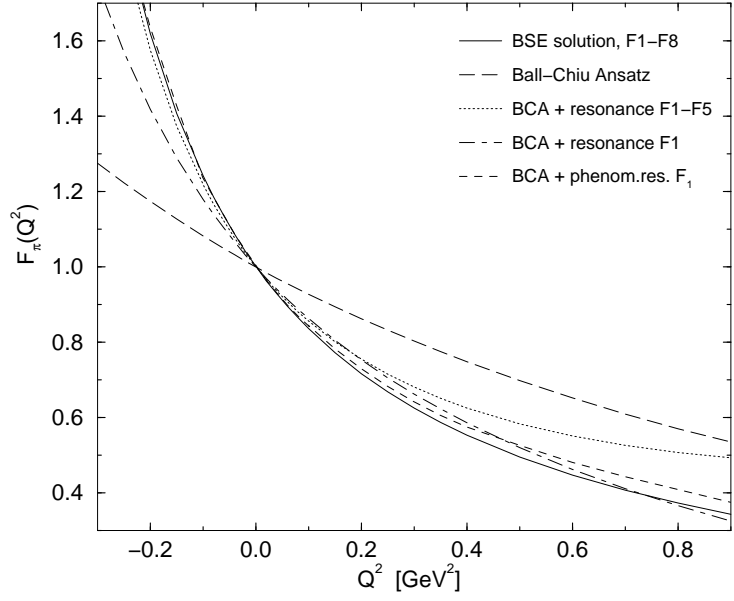,height=8.0cm} }
\caption{ 
The pion charge form factor $F_\pi(Q^2)$ as obtained from different
descriptions of the quark-photon vertex in terms of a resonant-improved
BC Ansatz, and compared to $F_\pi(Q^2)$ as obtained from the vertex BSE
solution.
\label{fig:piFFres} }
\end{figure}

The charge form factor resulting from two versions of the
resonant-improved BC Ansatz are shown in Fig.~\ref{fig:piFFres}.  One
version is based on the five dominant invariant amplitudes
${}^0F^V_i(q^2)$ of the $\rho$ meson, the other uses only the single
dominant amplitude ${}^0F^V_1(q^2)$; physical normalization is imposed
in each case.  The employed values of $m_\rho$ and $f_\rho$ in each case
are those obtained consistently from the homogeneous BSE in the present
model~\cite{MT99}, see Table~\ref{rhomf}.  It is seen from
Fig.~\ref{fig:piFFres} that both forms of the resonant-improved BC
Ansatz simulate the behavior of $F_\pi(Q^2)$ produced by the vertex
BSE solution quite well.

The obtained values of $r_\pi$ are shown in Table~\ref{piradii} for the
approximations discussed above.  Also shown there is the charge radius
produced by using the two dominant amplitudes $F_1$ and $F_5$ for the
resonant addition to the BC Ansatz.  Compared to \mbox{$r_{\pi,{\rm
BC}}^2 = 0.18~{\rm fm}^2$}, all versions of a resonant-improved BC
Ansatz provide results roughly a factor of two higher and in
significantly better agreement with the result from the vertex BSE
solution and the experimental data.  When the five dominant $\rho$
amplitudes are used, the resonant-improved BC Ansatz reproduces the
vertex BSE result for $r_\pi$ to within 5\%.  The error increases to
10\% if only the single dominant vector meson covariant is used for the
resonant term of the vertex; part of the reason for this is due to the
increase in the vector meson mass~\cite{MT99}: with $F_1$ only,
$m^2_\rho = 0.766~{\rm GeV}^2$, which moves the pole 30\% further into
the timelike region.

Use of Eq.~(\ref{piffbc+}) allows $r_\pi^2$ to be characterized as a
resonant addition to the result of the BC Ansatz.  If we write the
difference between $r_\pi^2$ and the BC contribution $r_{\pi,{\rm
BC}}^2$ as
\begin{equation}
        r_\pi^2 - r_{\pi,{\rm BC}}^2 =
                \frac{6 \, g_{\rho \pi\pi}\, F_{V\pi\pi}(0)}
                {m_\rho^2 g_\rho}\;.
\label{rbc+}
\end{equation}
and compare this with Eq.~(\ref{piffvmd}), we see that $F_{V\pi\pi}(0)$
characterizes the necessary weakening of the VMD mechanism for $r_\pi^2$
to account for the distributed $\bar q q$ substructure.  The values of
$F_{V\pi\pi}(0)$ defined by Eq.~(\ref{rbc+}), using the experimental
values for $g_{\rho\pi\pi}$, $m_\rho$, and $g_\rho$, are also shown in
Table~\ref{piradii}.  The difference between the $r_\pi^2$ result from
the vertex BSE solution and $r_{\pi,{\rm BC}}^2$ represents about 60\%
of the VMD value for the charge radius, \mbox{$r_{\pi,{\rm VMD}}^2 =
0.48~{\rm fm}^2$}.  The result from the best resonant-improved BC Ansatz
(using the vector meson amplitudes $F_1-F_5$) shows that the resonant
term is contributing to $r_\pi^2$ at the level of 50\% of the VMD value.
This decreases somewhat when the description of the vector meson is
simplified.  One can attribute this weakening of the VMD mechanism at
\mbox{$Q^2=0$} to the fact that the photon couples to a distributed,
interacting $\bar q q$ correlation and a significant part of this is
already accounted for by the BC Ansatz for the vertex.  With the
remainder viewed as due to the $\rho$ resonance, its effect is
overestimated if, for example, the coupling to $\pi\pi$ is described by
the physical mass-shell value $g_{\rho\pi\pi}$.  The effective reduction
of the $g_{\rho\pi\pi}$ is evidently about 50\% in the present approach;
this supports the previous cruder estimate~\cite{PCT98Adl}.

In a more phenomenological approach, using parameterized quark
propagators fitted to pion observables including $r_\pi$ as was done in
Ref.~\cite{MR98}, one obtains a much better value for $r_\pi$ using a BC
Ansatz.  In such an approach the $Q^2$-dependence of the vertex is
parameterized via the quark propagator, and can thus give a fair
representation of the pion form factor. The charge radius obtained in
this way, \mbox{$r_\pi=0.55~{\rm fm}$}, is much closer to the
experimental value than the one in the present work using a BC Ansatz.
This is also an indication of the model-dependence and arbitrariness of
separating the contributions into resonant and non-resonant terms.

The utility of the BC Ansatz for studies of electromagnetic coupling to
hadrons is that it is completely specified in terms of the quark
self-energy amplitudes.  For similar reasons we seek to summarize the
main features of the BSE solution for the quark-photon vertex also in
terms of the quark self-energy amplitudes and a phenomenological
$\rho$-meson BS amplitude, based on the resonance formula
Eq.~(\ref{bcplus}).  For the $\rho$ BS amplitude we use a simple Ansatz
for the dominant amplitude only, \mbox{$F_1^V(q^2) \rightarrow
N_\rho/(1+q^4/\omega^4)$}, properly normalized and with $\omega =
0.66~{\rm GeV}$ to give a decay constant $f_\rho = 201~{\rm MeV}$,
similar to the parameterization used in Ref.~\cite{IKR99}.  In order to
construct an Ansatz which can be used both at small and at large $Q^2$,
we consider the following phenomenological form for $Q^2 \ge -m_\rho^2$
\begin{eqnarray}
\label{phenver}
   \Gamma_\mu(q;Q) &=& \Gamma_\mu^{\rm BC}(q;Q) -
        \gamma_\mu^T \; \frac{N_\rho}{1+q^4/\omega^4}
        \frac{f_\rho \; Q^2}{m_\rho(Q^2 + m_\rho^2)}\, 
        e^{-\alpha (Q^2 + m_\rho^2)}\; .
\end{eqnarray}
Since the correct UV asymptotic limit is provided by $\Gamma^{\rm
BC}_\mu$, the parameter $\alpha > 0$ provides for a $Q^2$ suppression of
the resonant term.  We fit $\alpha$ to $r_\pi$, using the experimental
values for $m_\rho$ and $f_\rho$, and find good agreement with $\alpha =
0.03~{\rm GeV}^{-2}$.  The form factor $F_\pi(Q^2)$ produced by this
phenomenological vertex is shown in Fig.~\ref{fig:piFFres}, and is seen
to provide an excellent fit to the vertex BSE solution, and hence
according to Fig.~\ref{fig:piFF}, also to the experimental data.
Extensions to a more realistic $\rho$ BS amplitude are straightforward
to implement in phenomenological model calculations.

\section{Summary and Outlook}

We have used a ladder truncation for the BSE for the quark-photon
vertex, in conjunction with a ladder-rainbow truncation for the quark
DSE.  Both the vector WTI for the quark-photon vertex and the
axial-vector WTI are preserved in this truncation.  This ensures both
current conservation and the existence of massless pseudoscalar mesons
if chiral symmetry is broken dynamically: pions are Goldstone
bosons~\cite{MR97}.  The details of the model were fixed in previous
work~\cite{MT99}.  It leads to dynamical chiral symmetry breaking and
confinement; furthermore, at large momenta, our effective interaction
reduces to the perturbative running coupling and thus preserves the
one-loop renormalization group behavior of QCD and reproduces
perturbative results in the ultraviolet region.  The model gives a good
description of the $\pi$, $\rho$, $K$, $K^\star$ and $\phi$ masses and
decay constants~\cite{MT99}.

Our numerical solution of the vertex BSE shows clearly the vector meson
pole in all eight transverse amplitudes.  At the photon momentum $Q=0$,
the solution agrees perfectly with the BC Ansatz, as required by the WTI
and gauge invariance.  Also at spacelike asymptotic momenta our
transverse solution agrees with the BC Ansatz: both go to the bare
vertex.  At small but nonzero $Q^2$ the BSE solution departs
significantly from the BC Ansatz: the BSE solution has a stronger
$Q^2$-dependence.  In the region $-m_\rho^2 < Q^2 < 0.2~{\rm GeV^2}$,
the quark-photon vertex can be described by a BC Ansatz plus a resonant
term.  However, there is no unique decomposition of the vertex into
resonant and non-resonant terms away from the pole, and the BSE solution
for the vertex is the appropriate representation containing both
aspects.

\begin{figure}
\centering{\
\epsfig{figure=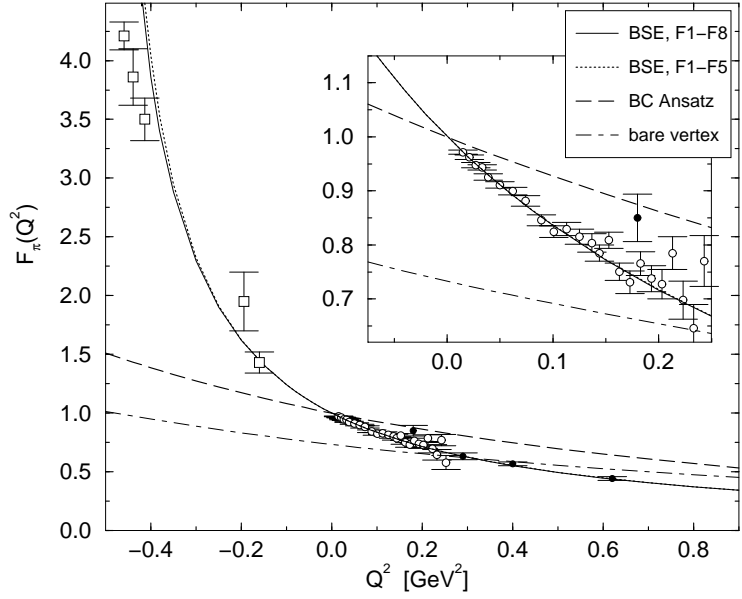,height=8.0cm} }
\caption{ 
The pion charge form factor $F_\pi(Q^2)$ as obtained from different
treatments of the quark-photon vertex.  The inset shows the $Q^2$ region
relevant for the charge radius.  The data correspond to $|F_\pi|$, taken
from Refs.~\protect\cite{B76} (circles), \protect\cite{B85} (squares),
and \protect\cite{A86} (dots).
\label{fig:piFF} }
\end{figure}

Subsequently, we have calculated the pion charge form factor in impulse
approximation.  The results using the BSE solution for the $\bar q q
\gamma$ vertex are in excellent agreement with the data, see
Fig.~\ref{fig:piFF}, and produce $r^2_\pi = 0.46~{\rm fm}^2$, {\em
without fine tuning the model parameters}: the parameters are completely
fixed in Ref.~\cite{MT99}.  Use of the BC Ansatz generally leads to a
charge radius which is too low: in the present model, the resulting
$r^2_\pi = 0.18~{\rm fm}^2$ is less than half the experimental value,
while the remainder can be described by a $\rho$ resonant term.  This
indicates that as much as half of $r_\pi^2$ can be attributed to a
reasonable extrapolation of the $\rho$ resonance mechanism.  On the
other hand, the strict VMD picture is too simple; about half of
$r_\pi^2$ arises from the non-resonant photon coupling to the quark
substructure of the pion.  One should however keep in mind that such a
separation in resonant and non-resonant contributions is ambiguous and
model-dependent.

The form factor $F_\pi(Q^2)$ exhibits a resonance peak at timelike
momenta $Q^2$ near $-m_\rho^2$, and our calculated $F_\pi(Q^2)$ does
indeed show such a peak.  Since in ladder truncation one does not
generate a width for the $\rho$ meson, we overshoot the data close to
the $\rho$-pole.  We expect that if we include the $\pi\pi$ mechanism
for the $\rho$ width in our formalism, we will have better agreement
with the data in the timelike region.  A detailed comparison close to
the $\rho$-pole will be postponed until we have included this effect in
our calculations.

Pion loops will not only generate a width for the $\rho$, they will also
give a direct contribution to the pion form factor and generate a
nonzero imaginary part for momenta $Q^2 < -4 m_\pi^2$.  Estimates are
that the pion charge radius increases by 10\% to 15\%~\cite{ABR95} due
to pion loops, which would lead to a charge radius which is too large in
the present model.  However, the pion loops will also affect the
calculated meson masses: estimates for the shift in the $\rho$ mass
vary between 2\% and 10\%~\cite{rhomass}, and the parameters in the
effective quark-antiquark coupling will have to be readjusted to
maintain agreement with the experimental data.  We hope to address these
question in future work.

With our BSE solution for the $\bar q q \gamma$ vertex we can now
investigate other hadron form factors.  First results
indicate~\cite{MT99Panic} that also the $\gamma^* \pi \gamma$ transition
form factor at low $Q^2$ is reasonably well described with the present
vertex BSE solution.  In the near future we will also apply this method
to the kaon form factor and to electromagnetic decays such as $\rho
\rightarrow \pi \gamma$.  Another interesting application is the nucleon form
factor, using models for the nucleon such as those being developed in
Refs.~\cite{BRSBF99,AAFOR99}.  The phenomenological resonance addition
to the BC Ansatz, Eq.~(\ref{phenver}), can be used as guidance for
phenomenological models to facilitate these and other hadron form factor
calculations, without having to solve the vertex BSE numerically.


\acknowledgements 
We acknowledge useful conversations and correspondence with
C.D.~Roberts, and D.~Jarecke.  This work was funded by the National
Science Foundation under grant No.~PHY97-22429, and benefited from the
resources of the National Energy Research Scientific Computing Center.

%
%

%
%
%
%

%
%
\begin{table}
\begin{center}
\begin{tabular}{l|dddd}
photon vertex           & $r_\pi$ & $r_\pi^2$& $r_\pi^2-r_{\pi,{\rm BC}}^2$ 
                                             & $F_{V\pi\pi}(0)$ \\ \hline
Expt.~\cite{A86}        & 0.663 $\pm$ 0.0006 & 0.44 $\pm$ 0.0012 
                                             &        &         \\ \hline
DSE $F_1-F_8$           & 0.678   &  0.460   &  0.278 & 0.58    \\
DSE $F_1-F_5$           & 0.677   &  0.459   &  0.277 & 0.58    \\
BC Ansatz (BCA)         & 0.426   &  0.182   &        &         \\
BCA + res. $F_1-F_5$    & 0.640   &  0.409   &  0.227 & 0.47    \\
BCA + res. ($F_1\&F_5$) & 0.625   &  0.390   &  0.208 & 0.44    \\
BCA + res. $F_1$        & 0.605   &  0.366   &  0.184 & 0.38    \\
BCA + phenom.res. $F_1$ & 0.68    &  0.46    &  0.28  & 0.58    \\
BCA of Ref.~\cite{MR98} & 0.55    &  0.30    &        &         
\end{tabular}
\caption{\label{piradii}
Results for the pion charge radius according to various treatments of
the quark-photon vertex. $F_{V\pi\pi}(0)$ roughly characterizes the
reduction in the $\pi\pi$ coupling strength for the vector $\bar q q$
correlation in the vector vertex compared to the on-shell coupling to
the $\rho$, $g_{\rho\pi\pi}$.}
\end{center}
\end{table}
\begin{table}
\begin{center}
\begin{tabular}{l|dd}
                & $m_\rho$ & $f_\rho$   \\ \hline
Expt.           &  0.770   &  0.216     \\ \hline
$F_1 - F_8$     &  0.742   &  0.207     \\
$F_1 - F_5$     &  0.730   &  0.201     \\
$F_1$ \& $F_5$  &  0.765   &  0.203     \\
$F_1 $          &  0.875   &  0.20       
\end{tabular}
\caption{\label{rhomf}
The vector meson mass $m_\rho$ and decay constant $f_\rho$ in GeV
corresponding to various truncations of the eight possible invariant
amplitudes of the $\rho$ BS amplitude~\protect\cite{MT99}.}
\end{center}
\end{table}
%
\end{document}